\documentclass{article}

\usepackage{PRIMEarxiv}
\usepackage{subcaption}
\usepackage[utf8]{inputenc} 
\usepackage[T1]{fontenc}    
\usepackage{hyperref}       
\usepackage{url}            
\usepackage{booktabs}       
\usepackage{amsfonts}       
\usepackage{nicefrac}       
\usepackage{microtype}      
\usepackage{lipsum}
\usepackage{fancyhdr}       
\usepackage{graphicx}       
\graphicspath{{media/}}     

\title{Prompt Migration: Stabilizing GenAI Applications with Evolving Large Language Models
}

\author{
  Shivani Tripathi, Pushpanjali Nema, Aditya Halder, Shi Qiao, Alekh Jindal \\
  Tursio Inc.\\
}

\begin{document}
\maketitle

\begin{abstract}
  Generative AI is transforming business applications by enabling natural language interfaces and intelligent automation. However, the underlying large language models (LLMs) are evolving rapidly and so prompting them consistently is a challenge. This leads to inconsistent and unpredictable application behavior, undermining the reliability that businesses require for mission-critical workflows. In this paper, we introduce the concept of prompt migration as a systematic approach to stabilizing GenAI applications amid changing LLMs. Using the Tursio enterprise search application as a case study, we analyze the impact of successive GPT model upgrades, detail our migration framework—including prompt redesign and a migration testbed, and demonstrate how these techniques restore application consistency. Our results show that structured prompt migration can fully recover the application reliability that was lost due to model drift. We conclude with practical lessons learned, emphasizing the need for prompt lifecycle management and robust testing to ensure dependable GenAI-powered business applications.
\end{abstract}
  
\section{Introduction}

Large Language Models (LLMs) have rapidly become the cornerstone of modern business applications, powering a diverse range of use cases from customer support chatbots and automated content generation to advanced data analytics and decision support systems. While on one hand organizations are increasingly relying on LLM-powered generative AI (GenAI) applications, on the other hand, the underlying models themselves are evolving at an unprecedented pace. For instance, in the first half of 2025, we already saw five major LLM releases from Open AI, and seven more from the likes of Google, Anthropic, DeepSeek and others. This rapid evolution poses significant challenges for businesses seeking to deploy stable GenAI applications that they can rely on. Ensuring that these GenAI applications remain robust and consistent, even as the foundational LLMs change, is critical for delivering dependable user experiences and safeguarding the business value.

Prompting is a key component of GenAI applications~\cite{Sahoo2025}, enabling developers to guide the model's behavior and responses. However, as LLMs evolve, the effectiveness of existing prompts can diminish, leading to inconsistent or degraded application behavior. Thus, we need a systematic approach to migrating existing prompts to newer LLMs, ensuring that applications can adapt to changes in the underlying models without sacrificing functionality or user experience. Prior work considered prompt evolution in the context of image generation, where prompts are refined iteratively to produce better results~\cite{Wong_2023}. This is different from chasing consistent results across multiple model versions, as the behavior of LLMs can vary significantly with each update. 

In this paper, we introduce the concept of {\it prompt migration} for stabilizing GenAI applications with evolving LLMs. We describe the evolution of Tursio~\cite{tursio} enterprise search application that was originally built on GPT-4-32k in 2023. Later, on June 6th, 2024, GPT-4-32k was deprecated with a shutdown date of June 6th, 2025 (i.e., a window of one year). Tursio migrated to GPT-4.5-preview, which was in turn deprecated on April 14th, 2025, with a shutdown date of July 14th, 2025 (i.e., a window of just three months). Currently, Tursio runs on GPT-4.1, and successive migrations lead to building a reliable testbed along with a set of practices for prompt migration. We discuss the Tursio war story and make the following key contributions:

\begin{itemize}
  \item We describe the Tursio enterprise search application and its dependence on Large Language Models (LLMs). (Section~\ref{sec:background})
  \item We present a detailed failure analysis of the Tursio application over successive GPT model versions, highlighting the challenges of prompt migration. (Section~\ref{sec:analysis})
  \item We show our prompt migration approach to evolve the Tursio application to GPT 4.1, and the corresponding testbed to handle future migrations. (Section~\ref{sec:migration})
  \item Finally, we discuss the lessons learned from a GenAI application perspective. (Section~\ref{sec:lessons})
\end{itemize}

\section{Background}
\label{sec:background}

\begin{figure*}[!t]
\centering
\includegraphics[width=\textwidth]{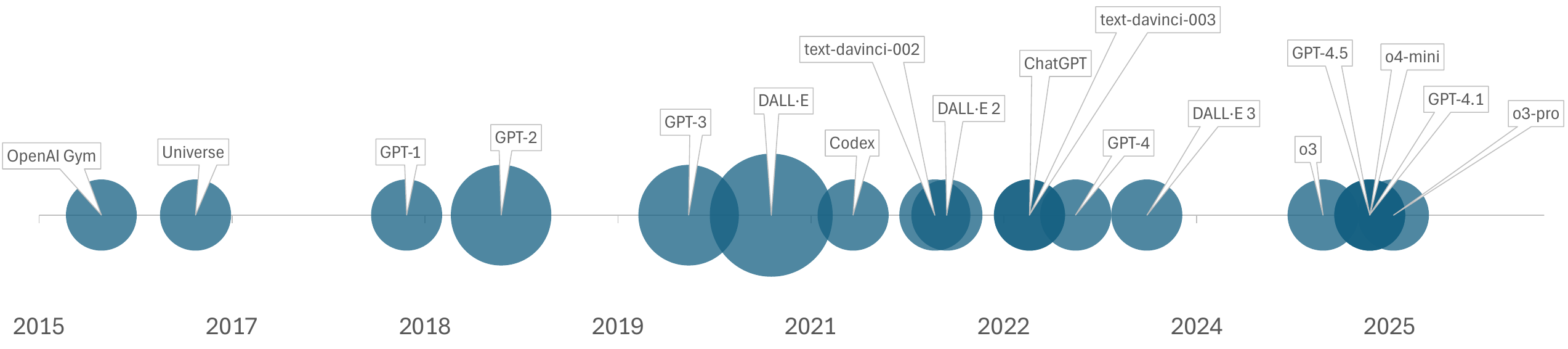}
\caption{Timeline of Open AI model releases.}
\label{fig:openai-models}
\end{figure*}

In this section, we discuss the rise of GenAI applications, the rapid evolution of LLMs that they have to cope with, and the challenges that Tursio enterprise search application faces in this context.

\subsection{Rise of GenAI Applications}

Generative AI is disrupting the traditional software-as-a-service (SaaS), with modern business applications shifting away from traditional SaaS-like CRUD operations and instead using generative language models to 
inject human-like reasoning. In fact, according to Satya Nadella, CEO of Microsoft, the SaaS as we know today is going to be dead~\cite{satya}. This is because unlike traditional SaaS applications that require months or years to build and often remain general-purpose tools, generative AI applications can be spun up in days or weeks while offering deep specialization tailored to specific business needs. Furthermore, generative AI is also simplifying the human-computer interface by replacing complex, multi-tabbed SaaS portals with intuitive natural language interactions, allowing users to simply describe what they need rather than navigate cumbersome menus and forms. 

Harvard Business Review notes that ``Generative AI is poised to revolutionize enterprise software, transforming traditional workflow systems into dynamic, goal-oriented environments.''~\cite{HBRGenAI}. No wonder industry adoption is accelerating rapidly, with GenAI budgets expected to grow $60\%$ from 2025 to 2027, expanding from $4.7\%$ to $7.6\%$ of total IT budgets~\cite{TechTargetGenAI}, while $53\%$ of C-level executives are now regularly using generative AI at work~\cite{MckinseyGenAI}. 

This trend reflects a broader transformation where businesses are moving from rigid, workflow-based software to intelligent, conversational applications that can understand intent, provide contextual responses, and execute complex tasks through simple human language commands, fundamentally changing how organizations interact with technology and dramatically reducing the time-to-value for enterprise software solutions.

\subsection{LLM Evolution}

LLMs are evolving rapidly and over the last few years the number of major LLM releases grew from $2$--$3$ in 2020 to $15$--$18$ in 2024, i.e., over 5x increase in new releases~\cite{LLMsWiki}. In 2025, $10+$ new LLM releases have already happened in first half of the year~\cite{LLMs2025}. To illustrate, Figure~\ref{fig:openai-models} shows the timeline of Open AI model releases, which is increasingly getting crowded in recent years and is representative of the overall trend in the industry. The rapid pace of LLM evolution presents both opportunities and challenges for developers and businesses. On one hand, new models often come with improved capabilities, better performance, and enhanced features. On the other hand, these changes can lead to inconsistencies in application behavior, requiring developers to adapt their prompts and workflows to align with the new model's characteristics.

LLM evolution has some similarities with schema evolution~\cite{10.14778/1453856.1453939} in that the frequency of change keeps on increasing while the applications are expected to continue operating on the new schemas. However, unlike schema evolution, where the changes are typically well-defined and versioned, LLM evolution is more dynamic and less predictable. Nevertheless, as GenAI applications start getting deployed more widely, we expect to see more attention on LLM evolution in the coming years.

\subsection{Tursio Enterprise Search}

Tursio is an enterprise search application that leverages Large Language Models (LLMs) to provide advanced search capabilities across various structured data sources, including SQL Server, Azure SQL, Microsoft fabric, Snowflake, Databricks, BigQuery, Teradata, and Cassandra. The underlying query engine in Tursio interprets user questions into well-formed operator trees. These trees consist of SQL operators such as select, project, filter, join, and aggregate, which are then executed against the data sources to retrieve relevant information. SQL operators can be further combined with ML and reasoning operations. Operator-based approach allows Tursio to handle complex queries and provide accurate results by leveraging the power of LLMs to understand natural language inputs and translate them into structured operators.  

Tursio was originally built on GPT-4-32k in 2023, which allowed it to handle large context windows and complex queries effectively. Over the years, Tursio has been deployed to over $100$ instances running thousands of queries per day. Each query triggers an analysis with $33$ LLM calls on average, including calls to basic foundational model, broader foundational model, and reasoning model. Together, these calls have a footprint of roughly $6{,}000$ input tokens and $1{,}000$ output tokens per query. Likewise, the training process incurs over $1{,}000$ LLM calls per run. Tursio clearly depends on LLMs, and so as they evolve, Tursio faces the challenge of maintaining consistent performance and reliability across newer model versions.

\section{Failure Analysis}
\label{sec:analysis}

In this section, we present a failure analysis of Tursio enterprise search application over successive GPT model versions. We first discuss the test results, then explain what failed, and finally dig into the differences between the models.

\subsection{Test Results}

We compare the effectiveness of Tursio's consistent prompts, referred here as the ``old prompts'', across multiple GPT versions (GPT-4-32k, GPT-4.1, GPT-4.5-preview). For example, the simplified version of the old prompt for extracting SQL filter operator from a fragment of the query text is as follows:

\begin{quote}
  \it
Extract SQL filters from the input. For column name, use underscores. Do not quote column names. Input: {fragment}. <Few-shot examples to guide model behavior>
\end{quote}
\vspace{0.2cm}

\begin{table}[!t]
\centering
\begin{tabular}{|l|l|}
\hline
Model & Tests Passed (\%) \\
\hline
GPT-4-32k & $100\%$ \\
GPT-4.1 & $98\%$ \\
GPT-4.5-preview & $97.3\%$ \\
\hline
\end{tabular}
\vspace{0.2cm}
\caption{Regression testing results for the Tursio enterprise search application.}
\vspace{-0.2cm}
\label{tab:tests}
\end{table}

Tursio's development environment safeguarded the ``old prompts'' via a large test suite. Unfortunately, the tests started failing with newer model versions. Table~\ref{tab:tests} shows the results over different model version. While perfectly stabilized for GPT-4-32k, the tests started failing with GPT-4.1 and GPT-4.5-preview, with $98\%$ and $97.3\%$ of the tests passing respectively. This indicates that the old prompts are not fully compatible with the newer models, leading to inconsistencies in the application behavior and breaking the CI/CD pipelines.

\subsection{Key Issues}

Let's look into the key issues that caused the above test failures. We categorize the issues into two main groups: common issues and model-specific issues.

\subsubsection*{Common Issues}
Some of the interpretations issues are common between 4.1 and 4.5-preview versions, including: 
\begin{itemize}
  \item Missing or incorrect interpretation of fragments that refer to \texttt{ORDER BY} or \texttt{GROUP BY} clauses.
  \item Incorrect column names (e.g., inferred non-existent ones that will obviously lead to later errors).
  \item Semantic misinterpretation (e.g., \texttt{TIMES\_LATE} misread as time dimension even though there was no such semantics).
  \item Ambiguous or absent filters when they should be inferred.
\end{itemize}

\subsubsection*{Issues Specific to 4.1}
Other issues are specific to GPT-4.1, including:
\begin{itemize}
  \item Requires more precise prompts to function correctly, e.g.,
  \subitem --- Misses implicit trends, such as failing to include \texttt{ORDER BY} for queries involving time (e.g., ``over time'').
  \subitem --- Fails to handle relative time expressions (e.g., ``last three months'') by omitting necessary \texttt{ORDER BY} clauses.
  \subitem --- Defaults to selecting whatever is mentioned in the question, rather than inferring from the context (e.g., ``analyze delinquency'').
  \subitem --- Does not perform its own inferencing unless explicitly instructed; needs clear instructions to extract information from the question or context.
  \item Mixes up column names and values, such as splitting values by underscores.
  \item Simplifies column names in a way that no longer matches the actual schema (e.g., \texttt{TIMES\_LATE}).
\end{itemize}

\subsubsection*{Issues Specific to 4.5}
Interestingly, GPT-4.5-preview faced all the above issues and have still more issues specific to it.
\begin{itemize} 
\item Incorrect output formats, leading to internal parser errors.
\item Printing info messages that could not be understood programmatically, e.g., ``could not find ..''.
\item Redundant operations in certain cases, e.g., introduce incorrect ORDER BY clauses.
\item Mixes up column name with operators to create incorrect columns.
\end{itemize} 

Clearly, the model behavior is drifting from GPT-4-32k to GPT-4.1 and then to GPT-4.5-preview, leading to the above issues.

\subsection{Model Differences}

The training objective in GPT-4.1 emphasizes stricter adherence to prompt instructions and prefers delegating uncertainties to external tools, reducing hallucinations compared to GPT-4. 
GPT-4.1 also supports a significantly larger context window, ranging from 100k up to one million tokens, whereas GPT-4's context window is limited to 2k to 8k tokens. GPT-4 was trained on approximately 2 trillion tokens; detailed training data size for GPT-4.1 is not publicly disclosed. Technical details for both models remain limited in public documentation.

These model differences can have significant impact. 
For instance, in terms of the objective function, both models are trained to maximize the likelihood of generating the next token given prior tokens, using cross-entropy loss and fine-tuning with Reinforcement Learning from Human Feedback (RLHF). However, GPT-4.1 places greater emphasis on following explicit instructions and minimizing unsupported inferences.
Likewise, the expanded context window in GPT-4.1 allows the model to process and reference much longer input sequences, enhancing its ability to generate contextually relevant and coherent responses.
For further technical details, see the GPT-4 technical report~\cite{openai2024gpt4technicalreport}.

\subsection{Observations from Release Notes}

The release notes for GPT-4.1 highlight its superior performance in coding, instruction following, and long-context comprehension. Of course, this comes with support for context windows of up to one million tokens. The evaluation metrics for this model have been shared publicly. In contrast, GPT-4.5-preview is described as an example of scaling unsupervised learning by increasing compute and data. It has been evaluated on academic benchmarks and outperforms the previous GPT-4o model. GPT-4.5-preview is recommended for general-purpose use due to its higher reasoning capabilities and demonstrated improvements in writing and design, as well as a higher ``EQ''. Additionally, it exhibits a lower hallucination rate compared to GPT-4o.

Furthermore, according to the release notes, GPT-4.5-preview demonstrates an improved ability to draw connections and generate creative insights compared to GPT-4.1. While detailed technical information about GPT-4.5-preview is limited, the release notes indicate that it was built on GPT-4o by scaling up pre-training data and model size, and by further generalizing capabilities using diverse data. Enhancements from the GPT-4o model, such as Supervised Fine-Tuning (SFT), reasoning with deliberative alignment, Chain-of-Thought reasoning, and Reinforcement Learning from Human Feedback (RLHF), are also included in GPT-4.5-preview.

Looking ahead, newer models are expected to deliver competitive performance on established benchmarks, demonstrate enhanced reasoning capabilities, and further reduce hallucinations. While these improvements are evident in recent releases, there are also ongoing rumors about the development of GPT-5, which may bring additional advancements.

\section{Prompt Migration}
\label{sec:migration}

In this section, we discuss our prompt migration approach to stabilize the Tursio enterprise search application with evolving LLMs. We first discuss the changes in model prompting guides, then present the new prompts added into our application, and finally describe the testbed that we built to support future migrations.

\subsection{Observations from Prompt Guide}

Let us now look into the prompting guides to understand what really changed in newer model versions.

It turns out that GPT-4.1 requires more explicit instructions. In fact, as per the GPT-4.1 prompting guide, \textit{``existing prompts optimized for other models may not immediately work with this model, because existing instructions are followed more closely and implicit rules are no longer being as strongly inferred''}~\cite{Open41Instruction}.

Interestingly, the prompting guide acknowledges the need for prompt migration, stating that \textit{``many typical best practices still apply to GPT-4.1, such as providing context examples, making instructions as specific and clear as possible, and inducing planning via prompting to maximize model intelligence. However, we expect that getting the most out of this model will require some prompt migration''}~\cite{Open41promptGuide}.

Specifically, the prompting guides recommends the following general prompt structure:
\begin{itemize}
  \item Role and Objective
  \item Instructions
  \subitem --- Sub-categories for more detailed instructions
  \item Reasoning Steps
  \item Output Format
  \item Examples
  \subitem --- Example 1
  \subitem --- Example 2
  \item Context
  \item Final instructions and prompt to think step by step
\end{itemize}

Reasoning prompts requires users to provide a reasoning strategy, the actual question, and any external context. Below is how the guide recommends structuring the reasoning prompts:

\begin{quote}
  \it

\# Reasoning Strategy

\vspace{0.2cm}
1. Query Analysis: Break down and analyze the query until you're confident about what it might be asking. Consider the provided context to help clarify any ambiguous or confusing information.

\vspace{0.2cm}
2. Context Analysis: Carefully select and analyze a large set of potentially relevant documents. Optimize for recall - it's okay if some are irrelevant, but the correct documents must be in this list, otherwise your final answer will be wrong. Analysis steps for each:
a. Analysis: An analysis of how it may or may not be relevant to answering the query.
b. Relevance rating: [high, medium, low, none]

\vspace{0.2cm}
3. Synthesis: summarize which documents are most relevant and why, including all documents with a relevance rating of medium or higher.

\vspace{0.2cm}
\# User Question
\{user\_question\}

\vspace{0.2cm}
\# External Context
\{external\_context\}

\vspace{0.2cm}
First, think carefully step by step about what documents are needed to answer the query, closely adhering to the provided Reasoning Strategy. Then, print out the TITLE and ID of each document. Then, format the IDs into a list ... 

\end{quote}

The prompting guides clearly reveal that the model behavior is changing, and so the prompts need to be adapted accordingly. The guides also recommend using few-shot examples to guide the model behavior, which is a common practice in prompt engineering. However, the examples need to be carefully crafted to align with the new model's capabilities and limitations.

\subsection{New Prompts}

To adapt to the changes in model behavior, we have introduced new prompts with specific instructions for the latest GPT versions. These prompts are structured to provide clear instructions, examples, and reasoning steps, as recommended by the prompting guides.

To illustrate, the new prompt for extracting SQL filters with explicit instructions from a fragment of the query text is as follows:

\begin{quote}
  \it
Extract SQL filters from the input question. The output must strictly follow this format: filters: [...]

\vspace{0.3cm}
Formatting Rules:
\begin{itemize}
  \item Enclose the filters in square brackets.
  \item Separate multiple operators with commas.
  \item Use underscores for column names instead of spaces.
  \item Do not quote column names.
  \item Infer implicit columns.
  \item The filter value should match the exact words or phrases used in the question, as closely as possible. Minor corrections for obvious typos are allowed.
  \item Refer to most similar question for inferring implicit filters.
  \item If there are no filters, return as filters: [].
\end{itemize}

\vspace{0.3cm}
Examples:
\begin{itemize}
  \item Question: List the demographics details for males after 2009.\\
        Filters: [Gender=`Male', Registration\_Date>`2009']
  \item Question: List demographics details for Spaniards between Jan 05, 2018
        and Aug 28, 2009.\\
        Filters: [Ethnicity=`Spaniard', Registration\_Date>=`2018-01-05', Registration\_Date<=`2009-08-28']
  \item Question: List demographics details for SSN not in 157549937 and 155485548 between 2018 and 2009\\
        Filters: [SSN!=`157549937', SSN!=`155485548', Registration\_Date>=`2018', Registration\_Date<=`2009']
\end{itemize}

\end{quote}

Using new prompts with GPT-4.1 resulted in $100\%$ tests results, i.e., full correction of the earlier issues with both precision and consistency.

\subsection{Migration Testbed}

Regression tests can only protect current workload but cannot guarantee the application readiness for the newer model versions.
To support future migrations, we have built a {\it migration testbed} that is more representative of the workload space and allows us to quickly adapt to new model versions. 
We have derived $110$ questions from hundreds of production workloads processed by Tursio into the migration testbed. We divide the testbed into three levels of complexity, easy, medium, and hard respectively, and describe each of them below.

\vspace{0.2cm}
\noindent{\bf Complexity: Easy.} We generate easy queries using the following preference order:
\begin{itemize}
  \item Queries having operators: Filter, GroupBy, Orderby, Limit.
  \item Queries having any of the 2 operators.
  \item	Queries having any one operator.
  \item	Resolve conflicting choices by preferring one with variations in columns and values.
  \item	Example: {\it Show top 3 city by lineitem count before Dec 2003.}
\end{itemize}

\vspace{0.2cm}
\noindent{\bf Complexity: Moderate.} We generate moderate queries using semantically similar variations of easy questions:
\begin{itemize}
  \item Prefer queries which we have failed previously while testing the SQL generation.
  \item Include samples from query generation cache, unit tests.
  \item	Example: {\it Show me visit details where there are ``no issue'' in detailed remarks}
\end{itemize}

\vspace{0.2cm}
\noindent{\bf Complexity: Hard.} We now generate queries that have implicit operations. Here is the processing needed for complexity level hard:
\begin{itemize}
  \item Included queries from the UTs which failed as we changed the models due to the presence of implicit columns, filters, aggregations, ordering.
  \item	Created queries by replacing words in the query with similar words.
  \item	Created queries which are follow more natural wording
  \item	Example: (1) List visits which required merchant as per detailed remarks.
  (2) List employees having more than 5 tasks over last two months.  
\end{itemize}

Prompt migration is non-trivial, and we faced several challenges at Tursio. 
Crafting meaningful queries that have good coverage requires good understanding of the underlying dataset, which is a time-taking process.
Furthermore, we could not include enough join queries beyond what is present in the existing unit tests since they depend on the application scenarios at hand. While our current migration testbed can be semi-generated for each of the three categories (easy, moderate, and hard), it still requires manual inspection for sanity and meaningfulness. Future work will look into making this process more automated.

\subsection{Putting It All Together}

We now describe our overall framework for prompt migration, a semi-assisted approach consisting of the following steps:

\begin{enumerate}
  \item Run the migration testbed on the new model version to identify the failing tasks.
  \item For failed tasks, update the corresponding prompts to the newer model version.
  \item Use the prompting guide as reference to learn the migration patterns and best practices. 
  \item Re-run the migration testbed and repeat steps 2--3 until all tests in the testbed pass.
  \item Run the regression tests to ensure the current workload is stable on the new model version; in case of failures, extend the migration testbed and repeat steps 1--4.
\end{enumerate}

\begin{figure}[!t]
\centering
\includegraphics[width=0.7\textwidth]{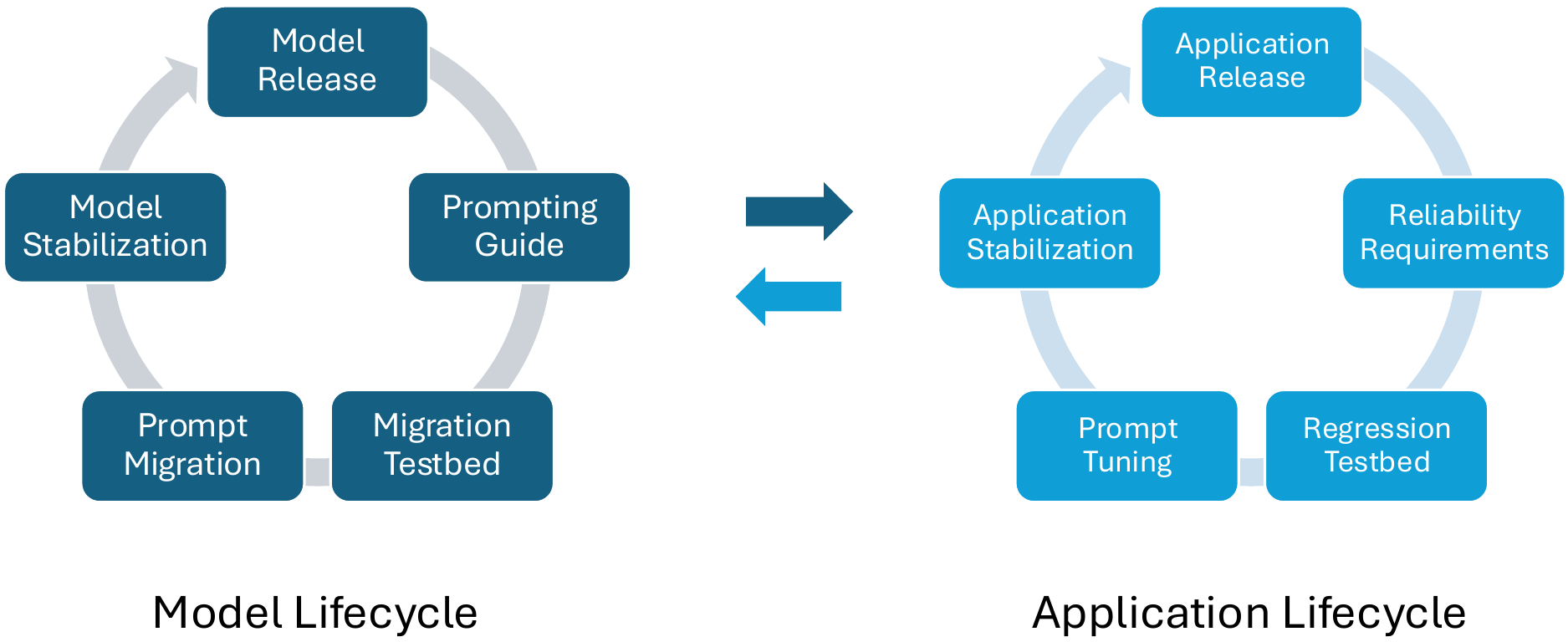}
\caption{Prompt lifecycle intertwined with model and application lifecycles.}
\label{fig:promot-lifecycle}
\end{figure}

Figure~\ref{fig:promot-lifecycle} shows the prompt lifecycle intertwined between model and application lifecycles --- new application releases add new prompts or tunes the existing ones, while new model releases require migrating the prompts for application stability. At Tursio, our first migration effort last couple of months end-to-end, but the above structured framework reduces it down to a couple of weeks. 

Note that steps 2 and 3 in our migration framework could be further accelerated using LLMs, i.e., provide the prompting guide and the failed prompt as input and ask the LLM to generate the modified new prompt. This will be part of our future work.

\section{Lessons Learned}
\label{sec:lessons}

We now discuss the key observations and lessons learned from the prompt migration experience of the Tursio application.

\vspace{0.2cm}
\noindent{\bf Flexible vs structured prompting.} 
We saw that GPT-4 performs best with flexible prompts, and it was able to understand implicit meaning.
On the other hand, GPT-4.1 and GPT-4.5-preview require more structured and explicit prompting to reach the full accuracy. Updating the prompts to these requirements significantly improved the results with GPT-4.1.
We also noticed that Langchain helps minimize hallucinations and increases adherence to instructions.

\vspace{0.2cm}
\noindent{\bf Example format matters.}
We found that the format in which the examples are provided matters a lot, e.g., plain text, JSON, XML, and so on.
Turns out that JSON and text are not that friendly and XML is better, as indicated in the OpenAI documentation, {\it ``when writing developer and user messages, you can help the model understand logical boundaries of your prompt and context data using a combination of Markdown formatting and XML tags''}~\cite{OpenAIFormatting}.

In fact, {\it ``XML is convenient to precisely wrap a section including start and end, add metadata to the tags for additional context, and enable nesting''}~\cite{Open41CookbookFormatting}. JSON on the other hand ends up being more verbose with more overheads due to character escaping.

\vspace{0.2cm}
\noindent{\bf Migration effort.}
Even though we came up with a representative migration testbed, the migration effort is still significant.
Moreover, newer models require more structured and verbose prompts, that grew from a few lines to a page in our migration effort. All this requires careful thinking and inspection from the developers, as well as tooling to cope up with the mushrooming models.

\vspace{0.2cm}
\noindent{\bf Prompt lifecycle.}
LLMs are changing how modern applications are being built. While prompt engineering was considered critical to these LLM apps so far, {\it prompt lifecycle management} is going to be another dimension. Prompts need to be continuously integrated with the newer model versions without breaking existing applications, thus making prompt migration a key part of the application development process.

Interestingly, our migration testbed approach decouples model exploration and evaluation from application development. These two processes typically involve different sets of people (AI engineers vs application engineers) anyways, and the migration testbed helps AI engineers to constantly experiment newer models. 

\vspace{0.2cm}
\noindent{\bf Prompt management skills.} 
Prompt engineering typical focuses on tuning the prompts to achieve the desired task. In contrast, {\it prompt management} is the operational part of it to migrate the tuned prompt to new model versions safely, i.e., without regressions --- an interesting new skill for people to learn as LLM-based applications find their way through every piece of our software stack.

\vspace{0.2cm}
\noindent{\bf LLM Intelligence.}
Fnally, we observed that LLMs are getting smarter in some ways, but dumber in others. For example, GPT-4.1 and GPT-4.5-preview require more explicit instructions and structured prompts, which is a step back from the implicit understanding that was possible with earlier models like GPT-4-32k. This raises an interesting question:
\begin{quote}
\it Is LLM a dumb master or a smart slave? 
\end{quote}

The rise of LLMs is opening up a brand-new world of possibilities, but it also raises broader questions about the nature of intelligence and reasoning in these models. For instance, more than $30\%$ of this paper has been generated with the help of LLMs, but they were also very hard to control in sticking to the point and on facts. Their ability to generate express amount of information is both a blessing and a bane --- ultimately, someone needs to be in control, e.g., via prompting, and that someone is us, the humans.

\section{Conclusion}
\label{sec:conclusion}

Generative AI is revolutionizing the way we build and interact with business applications, but the rapid evolution of Large Language Models (LLMs) poses significant challenges for developers. In this paper, we introduced the concept of prompt migration to stabilize GenAI applications with evolving LLMs. We presented the Tursio enterprise search application as a case study, detailing its evolution from GPT-4-32k to GPT-4.1 and GPT-4.5-preview, and the challenges faced during this transition.

We conducted a failure analysis of the Tursio application over successive GPT model versions, highlighting the issues that arose due to changes in model behavior and prompting requirements. We then described our prompt migration techniques, which involved updating prompts to align with the new model's capabilities and limitations, and building a migration testbed to support future migrations.

Finally, we discussed the lessons learned from this experience, including the importance of structured prompting, the need for prompt lifecycle management, and the evolving nature of LLM intelligence.
The future of human-LLM interface is going to be interesting and prompting will be a key part of it.

\section*{Acknowledgments}
We thank the Tursio engineering and products teams for deligently handling LLM migrations and coming up with a practical framework that led to this paper.

This paper was formatted using the \emph{Arxiv \& PRIME AI Style Template} by Moulay A. Akhloufi \cite{overleaf-template}, available on Overleaf under a CC BY 4.0 license.

\bibliographystyle{unsrt}  
\bibliography{paper}

\end{document}